\documentstyle[12pt]{article}

\def\br{\begin{thebibliography}{abc}}
\def\er{\end{thebibliography}}

\begin{document}

\baselineskip=24pt

\thispagestyle{empty}
\setcounter{page}{0}

\begin{flushright}
 IOP-HEP/030101 \\
 January 2003
\end{flushright}
{\large \centerline{\bf SOME REMARKS ON GRAVITY IN NONCOMMUTATIVE
SPACETIME} \centerline{\bf AND A NEW SOLUTION TO THE STRUCTURE
EQUATIONS }
 }
\vskip 1.5cm \centerline{ {\bf Nguyen Ai Viet} \footnote{\noindent
E-mail: nguyenaiviet@iop.ncst.ac.vn}} \vskip .5cm
\centerline{Department of High-Energy Physics, Institute of
Physics,} \centerline{National Centre of Science and Technology,
Hanoi, Vietnam} \vskip 1.5cm
\begin{abstract}
  In this paper, starting from the common foundation of Connes'
noncommutative geometry ( NCG)\cite{Connes1, Connes2, Connes3,
CoLo}, various possible alternatives in the formulation of a
theory of gravity in noncommutative spacetime are discussed in
detail. The diversity in the final physical content of the theory
is shown to be the consequence of the arbitrariness in each
construction steps.
  As an alternative in the last step, when the structure equations
are to be solved, a minimal set of contraints on the torsion and
connection is found to determine all the geometric notions in
terms of metric. In the Connes-Lott model of noncommutative
spacetime, in order to keep the full spectrum of the discretized
Kaluza-Klein theory \cite{VW2}, it is necessary to include the
torsion in the generalized Einstein-Hilbert-Cartan action.
\end{abstract}
\hspace{2cm}

\newpage

\section { Introduction}

   More than ten years after A.Connes \cite { Connes1, Connes2,
CoLo} suggested the idea of using NCG in Physics, the contruction
of a physical theory of gravity in noncommutative spacetime has
not reached a completely satisfactory state. The basic ideas of
generalizing Riemannian Geometry in noncommutative context have
been presented in \cite {Connes3, Kast, COQUE, Madore}. It was
Chamseddine and Fr\"ohlich who initiated the first real effort to
construct an extended Einstein theory of gravity utilizing the
noncommutative geometric model of Connes-Lott's two- sheeted
spacetime. Following this pioneering paper, a great number of
other papers have appeared, containing aspects of this interesting
topic. However, in these works, the physical content of the final
action, where the physical fields make their appearance is not
unique.

  Many papers have agreed with the field content of the standard Einstein
metric field along with a Brans-Dicke scalar, which represents the
distance between pairs of points on the two separate sheets.
However, in several other works \cite {WODZICKI}, with a different
formulation of the action, the scalar field does not have a
kinetic term, so it would not be present in the final theory as a
physical field. Some alternative constructions suggest that the
theory might have a Kaluza-Klein spectrum, which contains massless
and massive tensor, vector and scalar fields in a harmonic
expansion in the discrete internal spacetime.

  The reason for this diversity is due to the fact that although
most papers have accepted Connes'construction as the common
foundation, a unique recipe to build the final physical theory is
still lacking. With Connes' spectral triple given, there is still
considerable arbitrariness at every stage of the construction of
the final action that leads to different final results.

   The purpose of this paper is to point out the alternative choices at
various stages and find a new minimal set of constraints on the
torsion and connection to complete the Cartan structure equations.
This set is ``minimal'' in the sense that it does not place any
further constraints on the metric structure. Finally, based on
some physical considerations, a new solution to the structure
equations is found leading to the final action, where both torsion
and curvature contribute a satisfactory action.

\section{The common foundation: Connes' formulation}

   There is extensive literature about Connes' general formalism of NCG.
Here, for the sake of completeness, we shall begin with a brief
review of the relevant ideas that are necessary for this paper.
\cite{ Connes1, Connes2}.

   In Connes' formulation the basic building block of noncommutative
geometry is the spectral triple $( {\cal A}, D, {\cal H} )$:

   The algebra ${\cal A}$ of the 0-forms replaces the functions in a
generalized Gelfand's construction \cite{Connes1}. The replacement
of the number fields by the algebra ${\cal A}$ leads to the
concept of ${\cal A}$-bimodule \cite{Connes1} as a generalization
of the vector space.

   The Dirac operator $D$ acts on $F \in {\cal A}$ giving a subset of
the ${\cal A}$-bimodule of 1-forms. Therefore, $DF$ can be
formally written as
\begin{equation}\label{DERI}
  DF =   \sum_M (DF^M) (DF)_M, ~~ M \in \alpha, F^M \in {\cal A}
\end{equation}
where $\alpha $ is a finite set. That is to say, the
Eqn.(\ref{DERI}) postulates the existence of a finite basis
$(DF^M)$ of the set $\Omega^1({\cal A})$ of 1-forms. In other
words, $\Omega^1({\cal A})$ has an algebraic structure of finite
projective ${\cal A}$-bimodule. The elements $(DF^M)$ are also
given by the Dirac operator acting on $F^M \in {\cal A}$.

   The non-commutativity of the theory does not necessarily come
from the non-commutativity of the algebra ${\cal A}$. For example,
the Connes-Lott two-sheeted noncommutative space-time model
\cite{CoLo} is essentially based on the commutative algebra
$C^\infty( C, {\cal M}) \bigoplus C^\infty( C, {\cal M})$, where
${\cal M}$ is the usual physical spacetime. It is the Dirac
operator as an outer automorphism that brings about the
noncommutativity. Outer automorphism is the property of an
operator, whose action on an element gives an element outside of
the initial domain of the elements.

   It is possible to extend the definition of $D$ onto the bimodule of
1-forms $\Omega^1({\cal A})$ and  define the product of 1-forms in
order to build the ${\cal A}$-bimodule $\Omega^2({\cal A})$ of
2-forms. This procedure can be repeated to construct the universal
algebra $\Omega^*({\cal A})\equiv \bigoplus_p \Omega^p({\cal A})$
of differential forms on the algebra ${\cal A}$ with the Dirac
operator $D$. An involutive operation on ${\cal A}$ can be
extended uniquely to the one on the algebra $\Omega^*({\cal A})$.

The universal envelope algebra $\Omega^*({\cal A})$ has a graded
structure with the Dirac operator $D$ that takes
\begin{equation}\label{GRAD}
\begin{array}{ccc}
  D &:&  \Omega^p  \longrightarrow \Omega^{p+1} \hskip 4cm \\
  D((DF_1)....(DF_p) F) &\equiv& (-1)^{-1} (DF_1)...(DF_p) (DF),
  ~~~\forall F, F_1,..., F_p \in \Omega^p({\cal A}),
\end{array}
\end{equation}
which implies
\begin{equation}\label{BASE}
\begin{array}{cccc}
D^2F & = & 0, & ~~\forall p, F \in \Omega^*({\cal A})\\
D(F_1 F_2) & = & (DF_1) F_2 + (-1)^{degF_1} F_1 (DF_2),& ~~\forall
F_i \in \Omega^*({\cal A})
\end{array}
\end{equation}

   The Hilbert space ${\cal H}$ is where the
elements of ${\cal A}$, differential forms and exterior derivative
act as operators. In the Connes-Lott's model, it is chosen as the
direct sum of the Hilbert spaces of left-handed and right-handed
spinors ${\cal H} = {\cal H}_L \bigoplus {\cal H}_R $.

  The representation of the operators corresponding to differential forms
on a given Hilbert space is realized by the graded, involution
preserving homomorphism $\pi$,
\begin{equation}\label{PIREP}
\begin{array}{ccc}
  \pi &:& \Omega^*({\cal A}) \longrightarrow {\cal L}({\cal
  H}) \\
   \pi_p((DF_1)...(DF_p)F) &=& \prod_{i=1}^p [D, \pi_0(F_i)]
   \pi_0(F),
\end{array}
\end{equation}
where ${\cal L}({\cal H})$ denotes the space of bounded operators
on the Hilber space ${\cal H}$ and $\pi_0$ is a representation of
the algebra ${\cal A}$ on ${\cal H}$. Henceforth, the symbol $D$
will denote  both the Dirac operator $D$ and its representation as
a by a self-adjoint operator. It is assumed that $D$ has a compact
resolvent, such that the commutator $[D,\pi_0(F) ]$ is also a
bounded operator $\forall F \in {\cal A}$ \cite{Connes1}.

It is possible to choose a finite basis $(DF^M)$ as in
Eqn.(\ref{DERI}) and represent it by $\Theta^M = \pi_1((DF^M))$.
Therefore, an arbitrary 1-form $\omega \in \Omega^1({\cal A})$ is
represented as
\begin{equation}\label{1-FORM}
U = \pi_1 (\omega) = \Theta^M U_M.
\end{equation}
The Dirac operator is represented in the ``general'' basis
$\Theta^M$ as
\begin{equation}\label{DIRA}
\begin{array}{ccc}
D & = & \Theta^M D_M \\
(DF)_M & \equiv & D_M F = [D, \pi_0(F_M)].
\end{array}
\end{equation}

   With the differential forms in hand, one can
follow the standard procedure to introduce a metric structure, the
connection 1-forms, the torsion and, the curvature 2-forms to
construct a theory of gravity. The Cartan's structure equations
can be formulated and solved with some additional constraints to
express all the geometric quantities in terms of the metric. The
physical Einstein-Hilbert-Cartan action can be built from the
curvature and torsion.

   The formal procedure is rather well defined. Nevertheless, in
given specific representation, there are still alternate steps
that lead to diverse physical contents of the final theory of
gravity. In the next section, we will follow the steps in
formulating the theory by and point out the alternatives.

\section { Alternatives in the theory construction }

\subsection{The choice of the involutive operation and subalgebra}

   In the Connes-Lott model \cite{CoLo}, which concerns mainly with
the Standard Model, the physical fields are not represented by the
most general 1-forms. Instead, only the hermitian 1-forms are
considered as physical. Therefore, the involutive operation as the
hermitian conjugate ${~}*$ is introduced in the bimodule of
1-forms. In such a construction, the 0-forms continue to be
complex, while only hermitian 1-forms are relevant for physical
purposes. The hermitian forms do not form a ${\cal A}$-bimodule,
but rather as a bimodule of a subalgebra of ${\cal A}$, which is
the algebra of real functions ${\cal B} = C^\infty(R, {\cal M})
\bigoplus C^\infty( R, {\cal M})$.

  It suggest that a subalgebra of ${\cal A}$ maybe sufficient
to construct a gauge theory with the same physical content.

   In the forthcoming paper \cite{CHIRAL}, it is shown that, the Standard
Model can be constructed in the context of NCG, if one chooses the
subalgebra ${\cal A}'\in {\cal A}$ of the 0-forms in the basic
spectral triple as follows
\begin{equation}\label{0FORM}
F = \left(
    \begin{array}{cc}
     f(x) & 0      \\
     0    & f^*(x)
\end{array}
\right), ~~~ f(x) \in C^\infty( C, {\cal M})
\end{equation}
where $f^*(x)$ is the complex conjugate of $f(x)$.

  The complex conjugate will be coincident with the involutive
operation ``$ {~}^\sim $'' used in a series of papers
\cite{VW1,VW2,VW3,CHIRAL},
\begin{equation}\label{INVO}
{\tilde F} = F^* =  \left(
      \begin{array}{cc}
       f^*(x) & 0 \\
       0 & f(x)
\end{array}
\right)
\end{equation}

  It is noting that the restriction to the subalgebra ${\cal A'}$
is dicted by physics. In fact, one can start with the algebra
${\cal A}$ with the involutive operation defined as follows
\begin{equation}\label{TILD}
F = \left(\begin{array}{cc}
  f_1(x) & 0 \\
  0      & f_2(x)
\end{array}
\right) ~~,~~ {\tilde F} = \left( \begin{array}{cc}
  f_2(x) & 0 \\
  0 & f_1(x)
\end{array} \right), f_1(x), f_2(x) \in C^\infty( C, {\cal M}).
\end{equation}
It is shown in \cite{CHIRAL} that in order to have correct kinetic
terms for all the fields, one must restrict all the 0-forms and
all coefficients of higher differential forms subjected to the
algebra ${\cal A}'$.  The universal envelope algebra
$\Omega^*({\cal A}')$ constructed with the algebra ${\cal A}'$ can
be used to represent the physical fields without any further
condition.

  The example of the involution ``${~}^\sim$'' has shown that, in
  general, it is possible to choose various alternative involution operations
to have the desirable physical contents suitable for a given
application. This alternative produces the same field contents for
the gauge theory as the Connes-Lott model with a proper definition
of the inner product.

\subsection{Equivalence Principle}
   The foundation of General Relativity in
noncommutative spacetime is based on the Equivalence Principle. As
traditionally stated by Einstein, this principle postulates the
existence of the general and the locally orthonormal frames, which
can be transformed into each other by a local orthonormal
invertible transformation. The general frame consists of any
finite basis of the module of generalized 1-forms as formulated in
Sect.2. The exterior of derivatives of forms are defined in this
frame. Therefore, the differential calculus is always carried out
in this frame and the results will be transformed into other
frames if necessary. On the other hand, the local orthonormal
frame has the advantage in the algebraic calculations of forms at
the same location. Therefore, it is the most convenient basis for
the formulation of the structure equations for the connection,
torsion and curvature.

   In general, let us denote the basis of the general frame as $ \Theta^M $ and the
basis of the local orthonormal frame as $\Theta^A$. The
generalized vielbein is the spacetime dependent transformation:
\begin{equation}
\label{EQP}
\begin{array}{ccc}
\Theta^M & = & \Theta^A E^M_A(x) \\
\Theta^A & = & \Theta^M E^A_M(x),
\end{array}
\end{equation}
where the 0-forms $E^A_M(x)$ and $ E^M_A(x)$ are inverses of each
other,
\begin{equation}
\label{INVE}
\begin{array}{ccc}
  E^M_A(x) E^A_N(x) &=& \delta^M_N \\
  E^A_M(x) E^M_B(x) &=& \delta^A_B.
\end{array}
\end{equation}
  Since the derivatives are defined only in the ``general'' frame, the
Dirac operator $D$ is represented in the local orthonormal basis
as
\begin{equation}
D = \Theta^A E^M_A(x) D_M.
\end{equation}
  Indeed, the metric structure is encoded in the Dirac operator
via the presence of the vielbein $E^M_A(x)$. The structure of
$E^M_A(x)$ is where different choices are possible. In the most
general case, each 0-form $E^A_M$ contains a pair of the usual
functions.

  For the sake of definiteness, let us take the example in the Connes-Lott's
two-sheeted spacetime model, which is generally referred to by
most of the authors. In this model, the 0-forms and therefore the
vielbeins $E^A_M(x)$ and $E^M_A(x)$ where $ M = \mu, 5 $ and $A =
a, {\dot 5}$ are represented as $2 \times 2 $ matrices \cite{CHAM,
LVW, VW1, VW2, CHIRAL}. Specially, in the most general form, the
vielbeins are given as
$$
\begin{array}{c}
     E^\mu_a(x)
    \end{array} \equiv
\left(\begin{array}{cc}
    e^\mu_{1 a}(x) &  0 \\
     0    & e^\mu_{2 a}(x)
    \end{array}\right) ,\hspace{.5cm}
\begin{array}{c}
    E^\mu_{\dot 5} (x)
    \end{array} \equiv
\begin{array}{c}
     0
    \end{array},
$$
$$
\begin{array}{c}
    E^{5}_a(x)
    \end{array} \equiv -
\left(\begin{array}{cc}
    a_{1a} &  0 \\
     0    & a_{2a}
    \end{array}\right) \equiv
\begin{array}{c}
    - A_a(x) = - E^\mu_a A_\mu
    \end{array} ,
$$
\begin{equation} \label{VIEL}
\begin{array}{c}
    E^{5}_{\dot 5}(x)
    \end{array} \equiv
\left(\begin{array}{cc}
    \phi_1^{-1}(x) &  0 \\
    0     & \phi_2^{-1}(x)
    \end{array}\right) \equiv
\begin{array}{c}
    \Phi^{-1}(x)
    \end{array},
\end{equation}
where $e^\mu_{1,2a}(x)$ are two different vierbeins on the two
sheets of space-time. Similarly, $ a_{1,2}(x)$ and $\phi_{1,2}(x)$
are respectively vector and scalar fields. The vielbein
$E^\mu_{{\dot 5}}(x)$ can always be chosen as zero because of the
residue rotational arbitrariness.

    The vielbeins $E^M_A$ are
invertible as follows,
$$
\begin{array}{c}
     E^a_\mu(x)
    \end{array} \equiv
\left(\begin{array}{cc}
    e^a_{1 \mu}(x) &  0 \\
     0    & e^a_{2 \mu}(x)
    \end{array}\right) ,\hspace{.5cm}
\begin{array}{c}
    E^a_{5} (x)
    \end{array} \equiv
\begin{array}{c}
     0
    \end{array},
$$
\begin{equation} \label{VIEL-INV}
\begin{array}{c}
    E^{{\dot5}}_\mu(x)
    \end{array} \equiv
\begin{array}{c}
    A_\mu(x) \Phi(x)
    \end{array} , \hspace{.5cm}
\begin{array}{c}
    E_{5}^{\dot 5}(x)
    \end{array} \equiv
\left(\begin{array}{cc}
    \phi_1(x) &  0 \\
    0     & \phi_2(x)
    \end{array}\right) \equiv
\begin{array}{c}
    \Phi(x)
    \end{array}.
\end{equation}

  The Dirac operator is
\begin{equation}
\label{DIRA-MATR}
D = \left(
\begin{array}{cc}
\gamma^a e^\mu_{1a} \partial_\mu  &  -m \gamma^a e^\mu_{1a} a_{1\mu} + m \gamma^5 \phi_1 \\
\gamma^a e^\mu_{2a} a_{2\mu} -m \gamma^5 \phi_2 & \gamma^a
e^\mu_{2a}\partial_\mu
\end{array}\right)
\end{equation}

  In \cite{LVW}, by imposing the constraints $e^\mu_{1a} =
e^\mu_{2a}, a_{1\mu} = a_{2\mu}, \phi_1 = \phi_2$, the zero-modes
of Kaluza-Klein theory are obtained. In the other investigations
\cite { CHAM, SITARZ, CHINA} the vector field is assumed to
vanish.

  At this point, one can raise the question whether these restrictions are
internally required by the theory or just an arbitrary choice.

  Since Eqns. (\ref{VIEL}), (\ref{VIEL-INV}) and (\ref{DIRA-MATR}) cannot give any
restriction, one must look for further possibilities.

  The metric is given by a definition of the inner product, which
is a sesquilinear functional
\begin{equation}\label{METRIC}
\begin{array}{ccc}
 <~,~>~~~~ & : & ~~~~\Omega^1 ({\cal A}) X \Omega^1({\cal A})
 \longrightarrow {\cal A} \\
 < U F, V G > & = & {\tilde F} < U, V> G.
\end{array}
\end{equation}
In particular, one obtains
\begin{equation}\label{GMN}
 {\cal G}^{MN}(x) = < \Theta^M , \Theta^N > = {\tilde E}^M_A(x)
 \eta^{AB} E^N_B(x),
\end{equation}
directly from the othonormality of the frame $\Theta^A$ in the
following form
\begin{equation}\label{GAB}
 {\cal G}^{AB}(x) = < \Theta^A , \Theta^B > \equiv  \eta^{AB} =
 {\tilde E}^A_M(x) {\cal G}^{MN} E^B_M(x),
\end{equation}

From Eqn.(\ref{INVE}), that defines the inverse vielbein, one can
see that the equations (\ref{GMN}) and (\ref{GAB}) are in fact
equivalent. As these equations define the metric tensor in terms
of vielbein, one can conclude that they do not give any further
restriction on the vielbein.

   In \cite{CHINA} an argument is  made to show that the
vector field is related to a gauge transformation that emerges as
an internal shift in an analogy with the internal circle of the
ordinary Kaluza-Klein theory. However, the internal space of this
model consists of only two-points. Hence, one cannot speak about
the $U(1)$ gauge transformation in relation to the shift in the
internal circle. The gauge invariance must be guaranteed by a
different motivation.

   As proven in \cite{VW1}, the ``zero-mode only" constraint of
\cite{LVW} is a special case of the torsion free condition when
one solve the structure equation. There exist various ways to
impose conditions in order to solve the structure equations. Those
will be discussed later.

   One might also argue that in a $\Gamma$-representation, in which the
bases $\Theta^M$ and $\Theta^A$ are represented as $\Gamma^M$ and
$\Gamma^A$, the trace of these matrices might lead to a
restriction on the vector field.

   Let us consider this possibility, starting from the basis
\begin{equation}
\Gamma^a = \gamma^a \bigotimes {\bf 1}~~, \Gamma^{{\dot 5}} =
\gamma^5 \bigotimes {\bf \sigma},
\end{equation}
where
\begin{equation}\label{1sig}
\begin{array} {ccc}
   {\bf 1} & = & \left( \begin{array}{cc}
                           1  & 0 \\
                           0  & 1
                        \end{array}
                  \right) \\
   {\bf \sigma} & = & \left ( \begin{array}{cc}
                           0  & -1 \\
                           1  &  0
                         \end{array}
                  \right)
\end{array}
\end{equation}

  In this choice of basis, the ``general'' $\Gamma^M$ matrices are
\begin{equation}\label{GamM}
\begin{array}{ccc}
\Gamma^\mu & = & \left(\begin{array}{cc}
  \gamma^a e^\mu_{1a}(x) & 0 \\
  0 & \gamma^a e^\mu_{2a}(x)
\end{array}
\right) \\
\Gamma^5 & = & \left(\begin{array}{cc}
  -\gamma^a e_{1a}^\mu a_{1\mu} & \gamma^5 \phi_2^{-1}(x) \\
  \gamma^5 \phi^{-1}_1(x) & -\gamma^a e^\mu_{2a} a_{2\mu}
\end{array}
\right).
\end{array}
\end{equation}
   All the trace formulae of the $\Gamma$ matrices are consistent
with the metric tensors defined in Eqns.(\ref{GMN}) and
(\ref{GAB}). For example, the trace
\begin{equation}
\begin{array}{ccc}
{\cal G}^{\mu 5} & = &  Tr ( \Gamma^\mu \Gamma^5 ) ={\tilde
E}^\mu_A \eta^{AB} E^5_B = {\tilde E}^\mu_a \eta^{ab} E^5_b\\
& = & \left(\begin{array}{cc}
 - e^\mu_{1a} \eta^{ab} e^\nu_{1b} a_{1\mu}  & 0 \\
  0 & - e^\mu_{2a} \eta^{ab} e^\nu_{1b} a_{2\nu}
\end{array}
\right)
\end{array}
\end{equation}
gives the vanishing vector field only if one requires that $G^{\mu
5}$ vanishes. However, this happens only in specific cases such as
in the flat spacetime. In the general frame, in fact, one can
always choose the basis so that the metric ${\cal G}^{\mu 5} = 0$.
However, with such a choice, the tensor ${\cal G}^{AB}$ will not
be constant any more.

  The last possibility to have a constraint on the vielbein is to
see whether all its components have kinetic terms in the final
action. Without kinetic terms, these fields will not survive as
physical fields. If a field content in parallel with the
Kaluza-Klein spectrum is desirable, it will be shown that, one can
choose a set of minimal conditions in such a way that maximal
number of fields will survive.

\subsection{Alternative definitions of two-forms}

   In order to formulate the torsion and curvature in the structure
equations, one must define the ${\cal A}$-bimodule of 2-forms.
This module should not contain the ``junk forms''( the
non-vanishing forms that are differentials of the forms that are
indentical to zero in the $\pi$ representation). As an
illustration of a ``junk form'' in the usual differential
geometry, let us consider the 1-form $\omega = df.f - f.df$, which
is identical to zero. However, the ``junk form'' $d\omega =
-2df.df $ is non-vanishing. To eliminate these ``junk forms'', one
defines wedge products and replaces the ordinary product with the
wedge product to construct 2-forms as products or exterior
derivative of 1-forms.

   In NCG, one can also follow the technique of using auxialary
fields \cite { Connes1,COQUE} to find the general form of the
2-forms, which are not ``junk forms''. On the other hand, in the
case of Connes-Lott model, one can successfully define the wedge
product, which in fact eliminates the ``junk forms''
\cite{VW2,LVW,VW1,V1,V2}.

   The definition of the wedge product is not unique. In fact, one
can choose a wedge product that is fully anti-symmetric as in
\cite{VW2,LVW,VW1} for the theories of pure gravity . However, in
order to have a quartic Higgs potential for the gauge fields one
must choose a wedge product, where $ \Gamma^5 \wedge \Gamma^5 $ is
not zero. This arbirariness might result in additional terms in
the final action.

   Alternatively, in \cite{V1,V2} the wedge product is chosen as being fully
anti-symmetric, but one can still produce a quartic Higgs
potential, since the components of forms related to the internal
space is characterized by two complex quantities, which are
conjugates of each other, instead of one.

\subsection { Alternative sets of constraints}

  The torsion and curvature of Riemannian geometry can be
generalized via the structure equations in the local orthonormal
frame as follows
\begin{equation}\label{STRUCT1}
T^A = D\Theta^A - \Theta^B \Omega^A_{~~B}
\end{equation}
and
\begin{equation}\label{STRUCT2}
R^A_{~~B} = D\Omega^A_{~~B} + \Omega^A_{~~C} \Omega^C_{~~B},
\end{equation}
where $\Omega^A_{~~B}$ are the connection 1-forms. The structure
equations are not sufficient to determine the connection, torsion
and curvature in terms of metric fields. Therefore, some
additional constraints on the torsion and connection must be
imposed.

  In ordinary Riemmanian geometry, the torsion free condition
together with the metric compatibility equation completely
determines the connection and curvature in terms of the metric
structrure. In NCG, the torsion  free condition $T^A = 0$ has been
shown to lead to the following restriction on the metric structure
\cite{VW1}
\begin{equation}\label{DILATON}
\begin{array}{ccc}
e^\mu_{1a}(x) & = & \beta (x) e^\mu_{2a}(x), \\
a_{1\mu}(x)  & = & \beta(x) a_{2\mu}(x), \\
\phi_1(x) & = & \beta(x) \phi_2(x),
\end{array}
\end{equation}
where $\beta(x)$ is a dilaton field with a hightly non-linear
potential. In the special case of this restriction, where
$\beta(x) \equiv 1 $ one obtains the theory in \cite {LVW}
containing only the massless modes of the truncated Kaluza-Klein
theory with the vierbein, vector and Brans-Dicke scalar fields.

  Since a theory with the maximal spectrum would be interesting
for broader applications, it has been our purpose to find a
minimal set of constraints, which does not impose any restriction
on the metric. In \cite{VW2}, the following set has been found
\begin{equation}\label{FULLS}
\begin{array}{ccc}
T_{abc} & = & T_{ab{\dot 5}} = T_{a{\dot 5}b} = 0 \\
Tr ( T_{{\dot 5}AB} ) & = & 0,
\end{array}
\end{equation}
together with the metric compatible condition
\begin{equation}\label{MCOMP1}
\begin{array}{ccc}
\Omega_{AB} & = & - \Omega_{BA} \\
Tr(\Omega_{AB}{\bf  r}) & = & 0.
\end{array}
\end{equation}
  The set of constraints (\ref{FULLS}) and (\ref{MCOMP1}) are direct
generalizations of the spacetime torsion free and the metric
compatibility conditions. The additional constraints look rather
unnatural. However, it is the first model with the spectrum, which
is consistent with the intuitive discretized Kaluza-Klein theory.
When the internal space is discretized to just two points, a
harmonic expansion in the internal dimension would give gravity,
vector and scalar fields in pairs with their massive excitations.
In fact, all the vielbein components in the equation (\ref{VIEL})
survive in the curvature with their appropriate kinetic terms.

   In the next section, we show that it is possible to find another
minimal set of constraints. This illustrates the arbitrariness in
choosing alternative constraints.

\subsection { A formula for final action}

   With a proper set of constraints on the connection 1-forms and
torsion 2-form, one can solve the first structure equation
(\ref{STRUCT1}) to express all the components of the connections
and torsion completely in terms of vielbeins. Then, the curvature
is determined in terms of the connections from the second
structure equation (\ref{STRUCT2}).

   With all the geometric notions in hand, there is still an
arbitrariness in the formula of the final action. One may choose
the Wodzicki residue, which gives no kinetic term for the scalar
field \cite{WODZICKI}. With the same vielbein, the Dixmier trace
formula, on the other hand, contains the kinetic term for the
scalar field \cite{CHAM,SITARZ}. Therefore, the decision, whether
the scalar field can exist or not in the theory, depends on the
choice of the definition of action. Perhaps, using the Wodzicki
residue, one might have to include more characteristics into the
action to retain the scalar field in the final theory.

   The solution to be found in this paper shows a similar
situation: the constributions from the curvature do not contain
kinetic terms for the vector and scalar fields. To retain these
fields in the theory, one must include the quadratic term of the
torsion.

   It is important to note that in the formula of action that
contains the inner product, the definition of the involutive
operation may also alter the action.

\section{New constraints and new solution }

  In order to obtain the ordinary gravity, any set of constraints
must be an extension of the ordinary metric compatibility and
torsion free conditions.

  Let us discuss the metric compatibility condition first. The
simplest way is to generalized the metric compatibility condition
as
\begin{equation}\label{MCOMP-NEW}
\Omega_{AB} = - \Omega_{BA}.
\end{equation}
  Our previous work \cite{VW2,LVW,VW1} uses the same form of the
metric compatibility together the hermitian conjungate and a
reality condition. In this work, the metric compatibility
condition is consistent with the involutive operation
``${~}^\sim$'', which is originally defined on 0-forms, is just
the identity on 1-forms.

  The torsion free condition $T_A = 0$ has been proven to lead to
restriction on the vielbein \cite{LVW, VW1}. Therefore, we should
look for a weaker condition. The most direct generalization of the
torsion free condition is
\begin{equation}\label{TFRE}
T_a = 0.
\end{equation}

To see whether the metric compatbility and torsion free conditions
in the equations (\ref{MCOMP-NEW}) and (\ref{TFRE}) are enough to
solve the first structure equation, let us write its components as

\begin{equation}
\begin{array}{ccc}
  T_{abc}  & = & (D\Gamma_a)_{bc} + {1 \over 2}( \Omega_{abc} - \Omega_{acb})  \\
  T_{a{\dot 5}b} & = & (D\Gamma_a)_{{\dot5}b} + {1 \over 2}(\Omega_{a{\dot 5}b} - \Omega_{ab{\dot 5}}) \\
  T_{{\dot 5}ab} & = & (D\Gamma_{\dot 5})_{ab} + {1 \over 2}(\Omega_{{\dot 5}ab} - \Omega_{{\dot 5}ba}) \\
  T_{{\dot 5}{\dot 5}b} & = & (D\Gamma_{\dot 5})_{{\dot 5}b} +
           {1 \over 2}(\Omega_{{\dot 5}{\dot 5}b} - \Omega_{{\dot 5}b{\dot 5}}) \\
  T_{ab{\dot 5}} & = & (D\Gamma_a)_{b{\dot 5}}+
            {1 \over 2}(\Omega_{ab{\dot 5}} - \Omega_{a{\dot 5}b}) \\
  T_{{\dot 5} a {\dot 5}} & = & (D\Gamma_{{\dot 5}})_{a{\dot 5}}+
            {1 \over 2}(\Omega_{{\dot 5}a{\dot 5}} - \Omega_{{\dot 5}{\dot 5}a}) \\
  T_{a{\dot 5} {\dot 5}} & = & (D\Gamma_a)_{{\dot 5}{\dot 5}}+
            {1 \over 2}(\Omega_{a{\dot 5}{\dot 5}} + \tilde \Omega_{a{\dot 5}{\dot 5}}) \\
  T_{{\dot 5}{\dot 5}{\dot 5}} & = & (D\Gamma_{\dot 5})_{{\dot
  5}{\dot 5}} + {1 \over 2} ( \Omega_{{\dot 5}{\dot 5}{\dot 5}} + \tilde \Omega_{{\dot 5}{\dot 5}{\dot
  5}}).
\end{array}
\end{equation}

 It is obvious that the conditions (\ref{MCOMP-NEW}) and
 (\ref{TFRE})are not enough. There are various way to choose an
 additional condition. The simplest condition we can find is

\begin{equation}\label{3COND}
\Omega_{AB{\dot 5}} = 0
\end{equation}

  With three conditions (\ref{MCOMP-NEW}), (\ref{TFRE}) and
(\ref{3COND}) all the non-vanishing components of the torsion and
connection are given by
\begin{equation}
\begin{array}{ccc}
  \Omega_{abc}  & = & -(D\Gamma_a)_{bc} + (D\Gamma_b)_{ac} - (D\Gamma_c)_{ba} \\
  \Omega_{a{\dot 5}b} & = & - 2 (D\Gamma_a)_{{\dot 5} b} \\
  \Omega_{{\dot 5}ab} & = & 2 (D\Gamma_a)_{{\dot 5} b} \\
  T_{{\dot 5}bc} & = & (D\Gamma_{\dot 5})_{bc} + (D\Gamma_b)_{{\dot 5} c} - (D\Gamma_c)_{{\dot 5} b} \\
  T_{{\dot 5}{\dot 5}b} & = & (D\Gamma_{\dot 5})_{{\dot 5}b} \\
  T_{{\dot 5}{\dot 5}{\dot 5}} & = & (D\Gamma_{\dot 5})_{{\dot
  5}{\dot 5}}.
\end{array}
\end{equation}

   Since the exterior derivative of $\Gamma^A$ can be calculated in
terms of vielbein, the torsion and connections are expressed
completely in terms of the metric. Therefore, the second structure
equation (\ref{STRUCT2}) determines the curvature also in term of
the metric. No further restriction on the vielbein is required.

  Now if all the fields in the vielbein (\ref{VIEL}) will have
kinetic terms in the final action, they will survive as physical
fields.

  The scalar curvature is given
\begin{equation}\label{SCUR}
  R = < \Gamma^A \wedge \Gamma^B, R_{AB} >
\end{equation}
  After explicit calculation of the scalar curvature, we find
that the kinetic terms for the vector and scalar fields are not in
the scalar curvature as they do in the other solution in
\cite{VW2}.

  To retain these fields in the theory, the contribution from the
torsion must be included as the inner product
\begin{equation}\label{TCON}
<T_A, T^A> = {\tilde T}_{ABC} < \Gamma^B \wedge \Gamma^C, \Gamma^D
\wedge \Gamma^E> T^A_{DE}.
\end{equation}
  Therefore the final Einstein-Hilbert-Cartan Lagrangian of our theory is
\begin{equation}
{\cal L} = { 1 \over 16 \pi G_N} R + { 1 \over G^2_T} <T_A, T^A>,
\end{equation}
where $G_N$ is the Newton gravitational constant, $G_T$ is a new
constant introduced by the torsion.

   The calculation shows that all the kinetic terms are present.
With the mass terms in the final form of the Lagrangian, we come
to the conclusion that in this version our theory of gravity in
the Connes-Lott spacetime model contains the ordinary Einstein
gravity, one massive non-linear tensor, one massive and one
massless vector, one massive and one Brans-Dicke scalar fields.

\section{Conclusion }

   In this paper, a new specific model of gravity in noncommutative sapcetime
is proposed. However, in each step, the possible alternatives are
pointed out to show how one can retain the desirable contents by a
proper choice. In the final result, the solution of this paper
requires to include the torsion in the final Lagrangian to retain
the full spectrum of the discretized Kaluza-Klein model.

   From mathematical point of view, this construction has some
interesting features, which deserve some discussion. As stated
previously in this paper, the starting algebra ${\cal A} =
C^\infty( C, {\cal M}) \bigoplus C^\infty( C {\cal M})$ is too
large for physical applications. The hermiticity condition on the
physical 1-forms can give the correct physical content for the
gauge theory. However, the set of the hermitian forms does not
closed as ${\cal A}$-bimodule. As we shall see, our model can be
based on a minimal algebra for physics.

   The kinetic terms of the vielbein components must have correct
signatures to be physical fields. The signature will be correct
with the following restrictions on the vielbein components in the
equation (\ref{VIEL})
\begin{equation}
\begin{array}{ccc}
E^\mu_a(x) & = & \left(
             \begin{array}{cc}
              e^\mu_{+a} + i e^\mu_{-a} & 0 \\
                0   & e^\mu_{+a} - i e^\mu_{-a}
             \end{array}
             \right) \\

A_\mu(x)  & = & \left(
              \begin{array}{cc}
              a_{+\mu} + i a_{-\mu} & 0 \\
              0 &    a_{+\mu} - i a_{-\mu}
              \end{array}
              \right) \\

\Phi(x) & = & \left(
              \begin{array}{cc}
              \phi_+ + i phi_- & 0 \\
               0 & \phi_+ - i \phi_-
               \end{array}
               \right),
\end{array}
\end{equation}

where $e^\mu_{\pm a}, a_{\pm \mu}$ and $\phi_\pm$ are physical
real fields.

  With this restriction, the theory can be constructed consistently
from the subalgebra ${\cal A}'$ and $\Omega^p({\cal A}')$ will be
${\cal A}'$-bimodule. With this algebra in the spectral triple,
there is no need for additional condition on the differential
forms that represent the physical fields.

   The involutive operation ``${~}^\sim$'' on the 0-forms determines the
form inner product. This choice , in turns, is due to the decision
that we make about what kind of theory one wants to obtain in the
limit two sheets of spacetime become identical. The inner product
defined with the operation ``${~}^\sim$'' on 0-forms gives kinetic
terms, which are diagonal in $e^\mu_{a\pm}$, $a_{\mu \pm}$ and
$\phi_\pm$. In the limit, where two sheets of spacetime are
identical, all the massive modes $e^\mu_{a-}$, $a_{\mu-}$ and
$\phi_-$ of this model vanish and one obtain the truncated
Kaluza-Klein spectrum.

\bigskip
\noindent {\bf Acknowledgment s.}

     This work is dedicated to the memory of Victor Isacovich Ogievetsky. In our last
meeting in Philadelphia in Summer 1995, my plan on noncommutative geometry has
received enthusiastic encouragements from him. It is regretful that the realization of the
plan is so late what he cannot see the results.
     Thanks are also due to K.C.Wali for reading the manuscript
 and collaboration in the research program.

\bigskip
{\bf Appendix}

\end{document}